%% file: main.tex
\documentclass[sigconf]{acmart}

\usepackage{ifthen}
\usepackage{xspace}
\usepackage{booktabs}
\usepackage{listings}
\usepackage{algorithmic}
\usepackage{graphicx}
\usepackage{textcomp}
\usepackage{hyperref}
\usepackage{soul}
\usepackage[most]{tcolorbox}
\usepackage{balance}

\AtBeginDocument{%
  }

\copyrightyear{2024}
\acmYear{2024}
\setcopyright{rightsretained}
\acmConference[ASE '24]{39th IEEE/ACM International Conference on Automated
Software Engineering }{October 27-November 1, 2024}{Sacramento, CA, USA}
\acmBooktitle{39th IEEE/ACM International Conference on Automated Software
Engineering (ASE '24), October 27-November 1, 2024, Sacramento, CA, USA}
\acmDOI{10.1145/3691620.3695017}
\acmISBN{979-8-4007-1248-7/24/10}

\setcopyright{none}
\renewcommand\footnotetextcopyrightpermission[1]{}
\acmISBN{978-1-4503-XXXX-X/18/06}

\begin{document}

\title[AI-driven Java Performance Testing: Balancing Result Quality with Testing Time]{AI-driven Java Performance Testing:\\ Balancing Result Quality with Testing Time}


\author{Luca Traini}
\email{luca.traini@univaq.it}
\orcid{0000-0003-3676-0645}
\affiliation{%
  \institution{University of L'Aquila}
  \city{}
  \country{Italy}
}

\author{Federico Di Menna}
\email{federico.dimenna@graduate.univaq.it}
\orcid{0009-0003-5834-6389}
\affiliation{%
  \institution{University of L'Aquila}
  \city{}
  \country{Italy}
}

\author{Vittorio Cortellessa}
\email{vittorio.cortellessa@univaq.it}
\orcid{0000-0002-4507-464X}
\affiliation{%
  \institution{University of L'Aquila}
  \city{}
  \country{Italy}
}

\renewcommand{\shortauthors}{Traini et al.}

\titlenote{
This article has been accepted for publication in \emph{The 39th IEEE/ACM International Conference on Automated Software Engineering (ASE ’24)}.\\ This version of the manuscript is a preprint and may differ from the final published version in terms of content and formatting.\\
Citation information: \href{https://doi.org/10.1145/3691620.3695017}{DOI 10.1145/3691620.3695017}
}

\begin{abstract}
  Performance testing aims at uncovering efficiency issues of software systems.
  In order to be both effective and practical, the design of a performance test must achieve a reasonable trade-off between result quality and testing time. This becomes particularly challenging in Java context, where the software undergoes a warm-up phase of execution, due to just-in-time compilation. During this phase, performance measurements are subject to severe fluctuations, which may adversely affect quality of performance test results.  
  Both practitioners and researchers have proposed approaches to mitigate this issue.
  Practitioners typically rely on a fixed number of iterated executions that are used to warm-up the software before starting to collect performance measurements (\emph{state-of-practice}).
  Researchers have developed techniques that can dynamically stop warm-up iterations at runtime (\emph{state-of-the-art}).
  However, these approaches often provide suboptimal estimates of the warm-up phase, resulting in either insufficient or excessive warm-up iterations, which may degrade result quality or increase testing time.
  There is still a lack of consensus on how to properly address this problem.
  Here, we propose and study an AI-based framework to dynamically halt warm-up iterations at runtime. Specifically, our framework leverages recent advances in AI for Time Series Classification (TSC) to predict the end of the warm-up phase during test execution. We conduct experiments by training three different TSC models on half a million of measurement segments obtained from JMH microbenchmark executions. We find that our framework significantly improves the accuracy of the warm-up estimates provided by \emph{state-of-practice} and \emph{state-of-the-art} methods. This higher estimation accuracy results in a net improvement in either result quality or testing time for up to +35.3\% of the microbenchmarks. Our study highlights that integrating AI to dynamically estimate the end of the warm-up phase can enhance the cost-effectiveness of Java performance testing.
  \end{abstract}

%



\maketitle

\definecolor{pblue}{rgb}{0.13,0.13,1}
\definecolor{pgreen}{rgb}{0,0.5,0}
\definecolor{pred}{rgb}{0.9,0,0}
\definecolor{pgrey}{rgb}{0.46,0.45,0.48}

\lstdefinestyle{benchstyle}{   
  language=Java,
  showspaces=false,
  showtabs=false,
  keepspaces=true,
  breaklines=true,
  showstringspaces=false,
  breakatwhitespace=true,
  commentstyle=\color{pgreen},
  keywordstyle=\color{pblue},
  stringstyle=\color{pred},
  upquote=true,
  basicstyle=\ttfamily\footnotesize,
  columns=fullflexible,
  frame=tb,
  aboveskip=1em,
  belowskip=1em,
  lineskip=3pt
}

\lstset{style=benchstyle}


\newcommand{\ie}{\emph{i.e.,}\xspace}
\newcommand{\eg}{\emph{e.g.,}\xspace}
\newcommand{\etc}{etc.\xspace}
\newcommand{\etal}{\emph{et~al.}\xspace}
\newcommand{\secref}[1]{Section~\ref{#1}\xspace}
\newcommand{\figref}[1]{Fig.~\ref{#1}\xspace}
\newcommand{\listref}[1]{Listing~\ref{#1}\xspace}
\newcommand{\tabref}[1]{Table~\ref{#1}\xspace}
\newcommand{\algoref}[1]{Algorithm~\ref{#1}\xspace}
\newcommand{\eqqref}[1]{Equation~(\ref{#1})\xspace}
\newcommand{\tool}[1]{{\sc #1}\xspace}
\newcommand{\vda}{$\hat{A}_{12}$\xspace}
\newcommand*\circled[1]{\tikz[baseline=(char.base)]{
  \node[shape=circle,draw,inner sep=1pt] (char) {#1};}}
\newcommand{\unknown}{\textcolor{red}{XXX}\xspace}
\newcommand{\mysubsubsection}[1]{\medskip\noindent\textbf{#1}.\xspace}
\newcommand{\subsubsubsection}[1]{\smallskip\noindent\underline{\textit{#1}}.\xspace}

\newboolean{showcomments}

\setboolean{showcomments}{true}

\ifthenelse{\boolean{showcomments}}
  {\newcommand{\nb}[2]{
    \fbox{\bfseries\sffamily\scriptsize#1}
    {\sf\small$\blacktriangleright$\textit{#2}$\blacktriangleleft$}
   }
  }
  {\newcommand{\nb}[2]{}
  }

\newcommand\FEDE[1]{\textcolor{cyan}{\nb{FEDERICO}{#1}}}
\newcommand\VIC[1]{\textcolor{purple}{\nb{VITTORIO}{#1}}}
\newcommand\LUCA[1]{\textcolor{teal}{\nb{LUCA}{#1}}}


\input{introduction}

\input{background}
\input{approach}
\input{design}
\input{results}
\input{threats}
\input{related}
\input{conclusion}


\begin{acks}
This work was supported by the Italian Government (Ministero dell’Università e della Ricerca, PRIN 2022 PNRR) under the project ``RECHARGE: Monitoring, Testing, and Characterization of Performance Regressions'' (cod. P2022SELA7). Additional support was provided by the ``ICSC – Centro Nazionale di Ricerca in High Performance Computing, Big Data and Quantum Computing'', funded by the European Union – NextGenerationEU. This work also received support from the research start-up project ``Enhancing Software Performance Testing using Artificial Intelligence'', funded by the University of L’Aquila.
\end{acks}

\balance
\bibliographystyle{ACM-Reference-Format}
\bibliography{references}

\end{document}

%% file: introduction.tex
\begin{figure}[h]
	\center
	\includegraphics[width=0.9\linewidth]{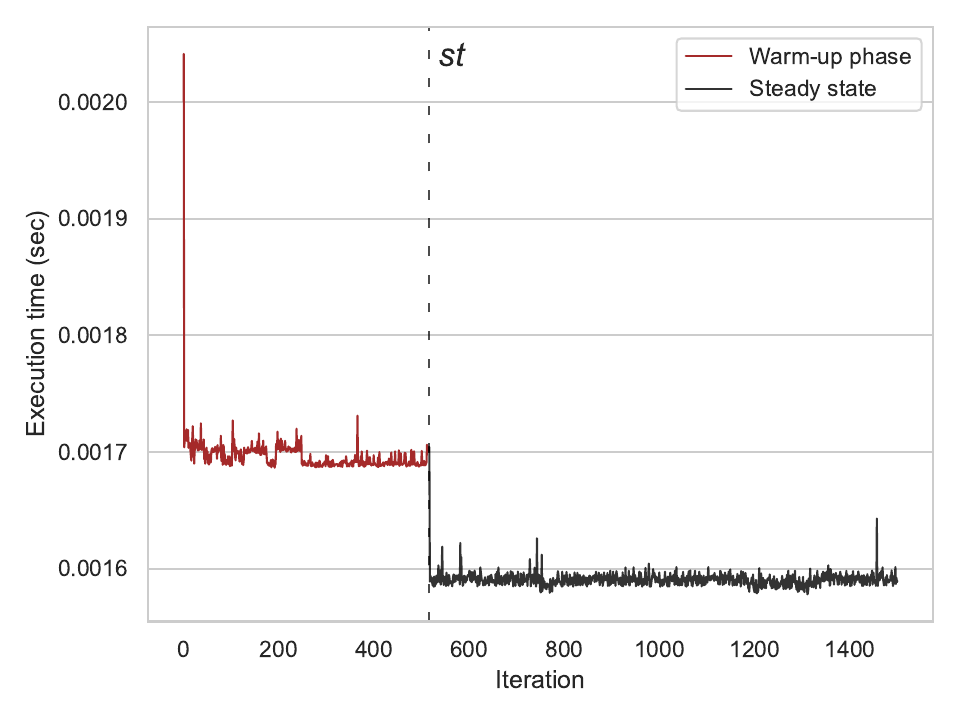}
	\caption{ Execution time of a Java microbenchmark (from the \emph{Roaring Bitmap} project) over consecutive iterations. The execution is characterized by an initial warm-up phase and a subsequent steady-state of performance. The grey dotted line indicates the iteration $st$ at which steady-state is attained.}
	\label{fig:ts-steadystate}
\end{figure}

\section{Introduction}\label{sec:intro}

Software performance is a critical non-functional aspect of software systems.
Technology organizations use performance testing to uncover performance bugs that might deteriorate software efficiency.
Nevertheless, performance testing requires careful design in order to be effective.
A significant challenge is to design an adequate number of execution repetitions that mitigate the variability of performance measurements~\cite{Maricq2018,Curtsinger2013,Mytkowicz2009}.
This typically involves balancing the need for quality of performance test results against the practical constraints of resources and testing time \cite{Kalibera2013a,Alghmadi2016,Chen2017,He2019a,Laaber2020a,AlGhamdi2023}.

Finding this balance becomes particularly difficult in the Java context, where software execution undergoes an initial warm-up phase due to just-in-time compilation \cite{Barrett2017a}.
During this phase, the Java Virtual Machine (JVM) performs a wide range of optimizations, leading to fluctuations in performance behavior (as shown in \figref{fig:ts-steadystate}) that may adversely affect result quality \cite{Georges2007a, Traini2022a}.
To mitigate this issue, it is common practice to conduct a number of ``\emph{warm-up iterations},'' with the sole goal of warming up the JVM, before starting to collect performance measurements.

An insufficient number of warm-up iterations might compromise the results quality, thus misleading performance evaluation.
On the other hand, unnecessary warm-up iterations increase testing time, thus hindering the adoption of performance testing in practice \cite{Jiang2015, Fagerstrom2016, Traini2022b}.
An accurate estimation of the warm-up phase is paramount to achieve cost-effective performance testing.

\begin{figure}[t]
	\center
	\includegraphics[width=1\linewidth]{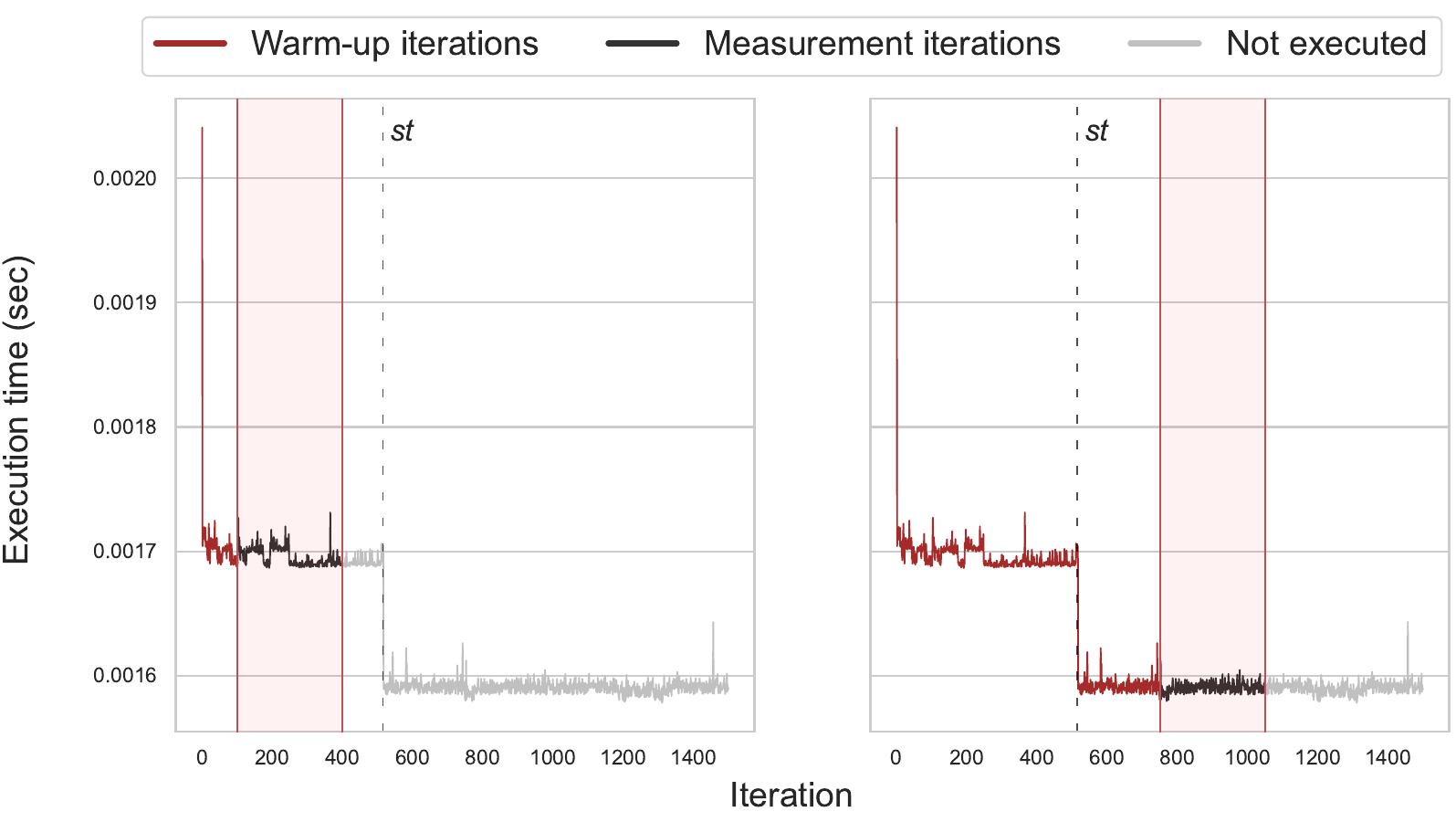}
	\caption{ 
	The left box shows an example of an insufficient number of warm-up iterations, which leads to the collection of measurements unrepresentative of the steady-state. The right plot depicts a scenario with an excessive number of warm-up iterations, which increases the testing time.}
	\label{fig:ts-warmup-estimation}
\end{figure}

Software engineers typically defines a fixed number of \emph{warm-up iterations} based on their domain expertise \cite{Laaber2020a}.
However, studies show that using a fixed number of iterations may misrepresent the warm-up phase \cite{Georges2007a, Laaber2020a, Traini2022a}.
In response to this, researchers developed state-of-art techniques that leverage statistical heuristics to dynamically stop the number of warm-up iterations at run-time \cite{Georges2007a, Laaber2020a}. 
Despite their advantages, these techniques are prone to significant inaccuracies that may either compromise result quality or increase testing time \cite{Traini2022a}.
It still remains unclear how to effectively stop warm-up iterations to ensure result quality in a timely manner.

In this paper, we propose an AI-based framework to dynamically stop warm-up iterations at run-time. Our framework utilizes Time Series Classification (TSC) models to predict whether a set of measurements is stable or not, and it exploits this capability to dynamically determine the end of the warm-up phase.
We integrate our framework with three state-of-the-art TSC models, and conduct experiments on JMH microbenchmarks, a form of performance testing commonly used in Java software \cite{Leitner2017}.
Results show that our framework noticeably improves the cost-effectiveness of performance testing.\\

The main contributions of our paper are summarized as follows:
\begin{itemize}
	\item We propose an AI-based framework that can integrate different TSC models to improve the cost-effectiveness of performance testing.
	\item We perform a comprehensive evaluation of our framework. The results show that, compared to both the state-of-practice and the state-of-the-art, our framework achieves a net improvement\textemdash in either result quality or testing time\textemdash in up to +27\% and +35.3\% of the microbenchmarks, respectively.
	\item We make publicly available the source code of our framework and the pretrained TSC models to aid future studies\footnote{Our replication package, including the dataset, source code and pre-trained TSC models, is publicly available at \url{https://doi.org/10.5281/zenodo.13749258}.}.
\end{itemize}

%% file: background.tex
\begin{center}
\begin{minipage}[t]{1\linewidth}
\begin{lstlisting}[language=Java, firstline=1, lastline=12, label={lst:JMH-example},
caption={Example of a JMH microbenchmark from the \emph{protostuff} Java library. The \texttt{@Warmup} annotation is used to define the number of warm-up iterations.}, captionpos=b
%,basicstyle=\fontsize{10}{12}\selectfont
]
@Warmup(iterations = 5)
@Measurement(iterations = 10)
public class StringSerializerBenchmark
{
    ...
    @Benchmark
    public byte[] builtInSerializer()
    {
        return s.getBytes("UTF-8");
    }
    ...
}
\end{lstlisting}
\end{minipage}
\end{center}

\section{Background}\label{sec:background}
In this section, we discuss the background related to our study, including Java microbenchmarking and Time Series Classification.

\subsection{Java Microbenchmarking}
Microbenchmarking is a type of unit-level performance testing, commonly used in Java context \cite{Leitner2017}.
A microbenchmark repeatedly executes a small portion of code, such as a Java method, while gathering measurements of its execution time.
Due to just-in-time compilation, microbenchmarks are subject to performance variability in the first phase of their execution,\footnote{
The JVM allows disabling JIT compilation to reduce performance variability. However, doing so would give an unrealistic representation of software performance, as JIT compilation is typically enabled in production environments \cite{oaks2014}.} also known as \emph{warm-up}.
During this phase, the JVM detects frequently executed loops or methods, and it dynamically compiles them into optimized machine code.
After the completion of the warm-up phase, the microbenchmark is said to be executing at a \emph{steady-state of performance} \cite{Barrett2017a}.

Software engineers employ \emph{warm-up iterations} to ensure that the microbenchmark achieves a steady-state before starting to collect measurements. 
As shown in Figure \ref{fig:ts-warmup-estimation}, executing an appropriate number of the warm-up iterations is paramount to achieve cost-effective performance testing.
Insufficient number of warm-up iterations may produce measurements that do not reflect the software’s true steady-state performance, thus detrimentally affecting result quality.
Conversely, excessive warm-up iterations lead to unnecessary executions that prolong the testing time.


\mysubsubsection{State-of-practice (SOP)}
Practitioners usually predefine a fixed number of warm-up iterations based on their domain expertise.
To configure warm-up iterations, they typically rely on Java Microbenchmark Harness (JMH) \cite{JMH}, the de-facto standard for building, configuring, and running Java microbenchmarks. 
JMH allows to configure the number of warm-up iterations directly in the microbenchmark's source code, using Java annotations, as shown in Listing \ref{lst:JMH-example}.
Nevertheless, prior work has shown that using a fixed number of iterations may often produce suboptimal estimates of the warm-up phase \cite{Georges2007a, Laaber2020a, Traini2022a}.

\mysubsubsection{State-of-the-art (SOTA)}
Researchers have developed techniques that can dynamically stop warm-up iterations at run-time. Georges \etal \cite{Georges2007a} introduced a first such approach, using a preset threshold on the coefficient of variation \cite{Everitt2010} to determine the end of the warm-up phase.
However, subsequent studies have identified Georges \etal's heuristic as both inaccurate \cite{Kalibera2013a} and unrealistic \cite{Laaber2020a, Traini2022a} in practical scenarios.
Another technique was recently proposed by Laaber \etal~\cite{Laaber2020a}. Similarly to Georges \etal's heuristic, it leverages statistical metrics to estimate the end of the warm-up phase. At each microbenchmark iteration, this technique checks whether adding more performance measurements are likely to change the distribution of performance measurements, and it uses this information to dynamically stop warm-up iterations at runtime. Laaber \etal provide three different variants of this technique, based on distinct statistical metrics, namely coefficient of variation (COV)~\cite{Everitt2010},\footnote{Both the approaches of Georges \etal \cite{Georges2007a} and Laaber \etal \cite{Laaber2020a} use the CV, but they apply it in different ways. Georges \etal use a fixed threshold on CV, whereas Laaber \etal verify whether the addition of more measurements changes the CV within a specified threshold.} relative confidence interval width~\cite{Davison1997,Kalibera2012a} (RCIW), and Kullback-Leibler divergence~\cite{Kullback1951, He2019a} (KLD). 
Prior work has shown that Laaber \etal's technique is more accurate than the SOP in estimating the end of the warm-up phase \cite{Traini2022a}.


\subsection{Time Series Classification}
Time Series Classification (TSC) involves assigning predefined labels to time-ordered sequences of data points based on their characteristics or patterns.
TSC problems arise in many domains, such as human activity recognition \cite{Chen2021a}, e-health \cite{Rajkomar2018}, natural disasters \cite{Arul2021}, and finance \cite{Majumdar2020}.
Due to its broad applicability, hundreds of TSC algorithms have been proposed over the last decades \cite{Middlehurst2023,Bagnall2017}.
These algorithms span various categories, including distance-based \cite{Lines2015}, dictionary-based \cite{Schafer2015}, convolutional-based \cite{Tan2022, Dempster2020a}, or deep learning approaches \cite{Fawaz2019}.
Recently, the latter two categories have demonstrated notable advancements, enabling them to attain state-of-the-art performance on established TSC benchmarks \cite{Middlehurst2023, Fawaz2019, Foumani2023, Tang2022a}, such as the UCR archive \cite{Hoang2018}.
In this work, we study the efficacy of three state-of-the-art TSC algorithms to dynamically stop warm-up iterations of microbenchmarks. Specifically, we investigate one convolutional-based algorithm, namely ROCKET \cite{Dempster2020a}, and two deep learning models, namely FCN \cite{Wang2017a} and OSCNN \cite{Tang2022a}.

%% file: approach.tex
\section{Methodology}\label{sec:approach}
In this section, we present the methodology of our AI-based framework. We first introduce an overview of the main phases of our framework and then discuss the details of each phase.

\begin{figure*}[htbp]
\center
\includegraphics[width=0.8\linewidth]{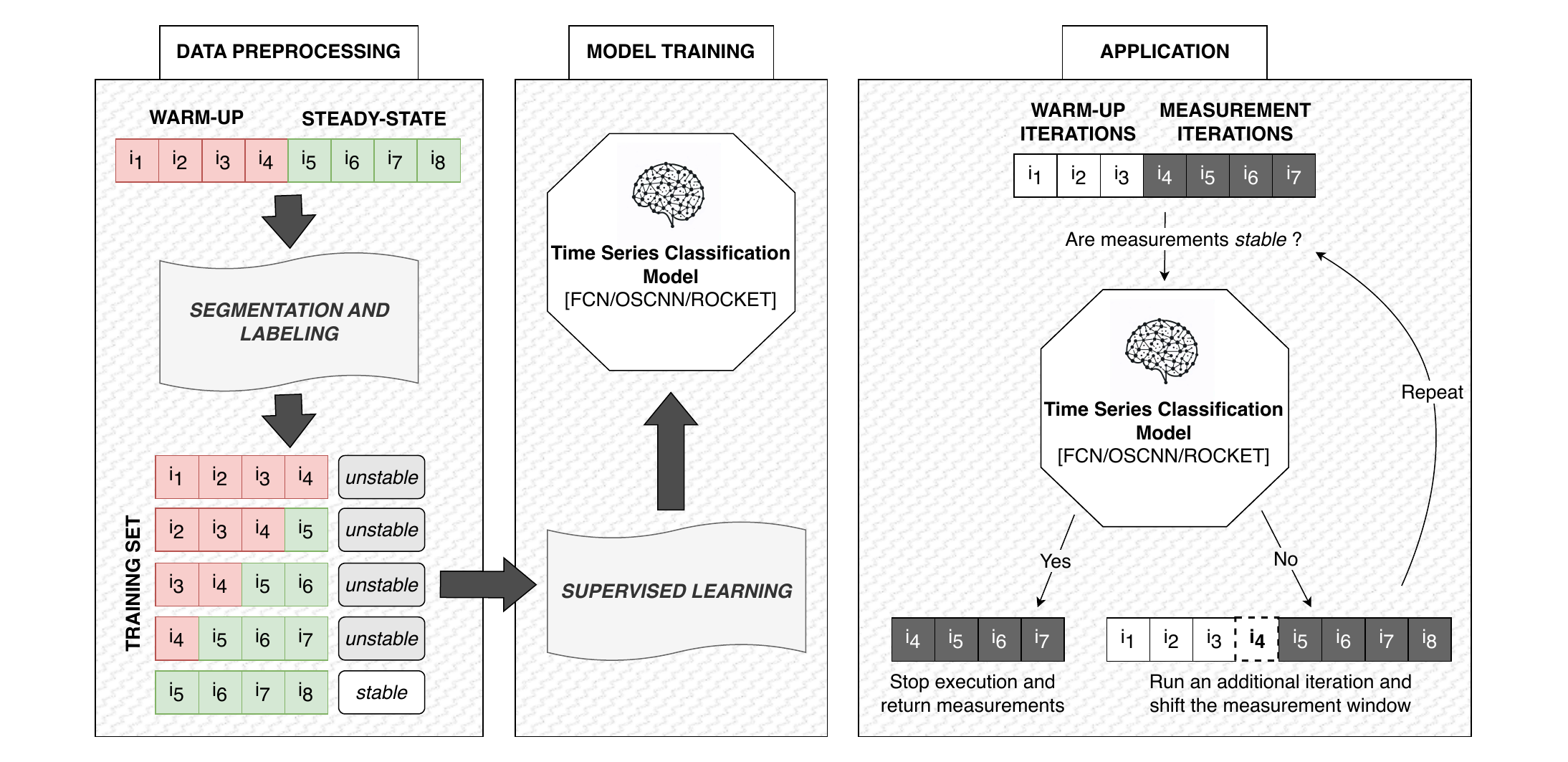}
\caption{ Overview of the main phases of our framework: \textbf{\textit{Data Preprocessing}}, \textbf{\textit{Model Training}} and \textbf{\textit{Application}}.
}
\label{fig:overview}
\end{figure*}

\subsection{Overview}
As shown in \figref{fig:overview}, our framework involves three main phases: \emph{Data Preprocessing}, \emph{Model Training}, and \emph{Application}. The first two phases are executed offline, the last one during performance test execution.

During the \textit{Data Preprocessing} phase, our framework begins by processing a time series of performance measurements with a known warm-up end. This process involves segmenting the time series and labeling each segment as \textit{stable} or \textit{unstable} based on the presence of warm-up measurements. 
The labeled segments are then used for training the Time Series Classification model (\textit{Model Training}).
Finally, the framework leverages the trained model at runtime to dynamically stop the number of warm-up iterations (\textit{Application}).

\subsection{Data Preprocessing}
This phase aims at preparing the dataset that will be used for supervised learning.
To achieve this, our framework starts from a time series of performance measurements, and an annotation that denotes the end of the warm-up phase.
The time series represents the observed execution time over sequential iterations of a JMH microbenchmark.
The annotation marks the specific iteration ($st$) in which the steady-state of performance is reached.

In this study, we rely on the notion of steady-state defined by Barrett \etal \cite{Barrett2017a}. This notion leverages change point detection, namely PELT \cite{Killick2012}, to determine statistically significant shifts in the time series, and to identify the end of the warm-up phase.
Barrett \etal's approach cannot be used at run-time to determine warm-up, since it requires to first run the microbenchmark for an unrealistically large number of iterations, and, subsequently determine the end of the warm-up phase through post-hoc analysis. However, we aim to leverage this approach to build a solid ground-truth for our AI-based framework.

To construct our dataset, we sample smaller (overlapping) segments of fixed size from the original time series, with each segment representing a contiguous block of performance measurements.
We then classify these segments with binary labels:

\begin{itemize}
    \item \emph{Stable}: The segment consists solely of measurements from the steady-state, meaning all measurements were taken after the steady-state $st$ was reached.
	\item \emph{Unstable}: The segment includes at least one measurement from the warm-up phase, indicating that some measurements were taken before reaching $st$.
\end{itemize}

\figref{fig:overview} illustrates a simplified representation of this process.

Our dataset construction involves multiple time series gathered from various microbenchmarks across different software systems. This process yields a heterogeneous dataset of measurement segments that capture diverse performance patterns from a range of software systems.

\subsection{Model training}
Our learning goal implies the binary classification of segments of performance measurements.
We leverage Time Series Classification algorithms to classify each segment as either \emph{stable} or \emph{unstable}.
Specifically, we study three different TSC algorithms, which we discuss below:
\medskip

\noindent\textbf{Fully Convolutional Network (FCN)} was proposed by Wang \etal \cite{Wang2017a}
for classifying univariate time series. This neural network architecture acts as feature extractor stacking three convolutional layers \cite{Fukushima1980}, each followed by a batch normalization layer \cite{Ioffe2015} and ReLU activation layer \cite{Fukushima1969}.
The features are then processed through a global average pooling layer \cite{Lin2014} and finally passed into a softmax classifier to obtain the final output label.
Prior work has shown that this architecture can achieve state-of-the-art performance in TSC \cite{Fawaz2019}.
\medskip

\noindent\textbf{Omni-Scale Convolutional Network (OSCNN)} introduces the Omni-Scale block \cite{Tang2022a},  wherein the kernel sizes for 1-dimensional convolutional neural networks (1D-CNNs) are automatically set through a simple and universal rule. 
This approach enables capturing an optimal receptive field size, that is a factor that significantly influences the performance of 1D-CNNs for TSC \cite{Cui2016}.
\medskip

\noindent\textbf{Random Convolutional Kernel Transform (ROCKET)} is a pipeline classifier \cite{Dempster2020a}.  It generates a large number of randomly parameterized convolutional kernels and uses these to transform the data through two pooling operations: max value and the proportion of positive values. These two features are concatenated into a feature vector for all kernels. The feature vectors are then used to train a Ridge Classifier \cite{Hilt1977} using cross-validation.
Unlike convolutional neural networks, ROCKET does not involve any hidden layers, thereby providing state-of-the-art performance with limited computational cost \cite{Tang2022a, Middlehurst2023, Dempster2020a}.

\subsection{Application}

Our framework employs TSC models to dynamically predict when the microbenchmark reaches a steady-state of performance.
As the microbenchmark execution progresses, the framework continuously evaluates the incoming performance measurements using the TSC model.
Once the model detects the achievement of the steady-state, the framework promptly halts the microbenchmark execution and returns the current set of measurements for performance evaluation.
\figref{fig:overview} presents a snapshot of this procedure using an illustrative scenario. In this scenario, the microbenchmark has completed 3 warm-up iterations, with the current measurement window ranging from the 4th to the 7th iteration. Before proceeding, our framework queries the TSC model, forwarding the current set of measurements to determine if they correspond to a steady-state of execution. If the model predicts that the measurements are \emph{stable}, the microbenchmark execution is halted, and these measurements are returned for performance evaluation. Conversely, if the model classifies them as \emph{unstable}, then the framework makes the test run another iteration, and it shifts the measurement window by one iteration. This process repeats at each iteration until the model identifies a \emph{stable} segment (or until a predefined maximum number of iterations is achieved).

The key insight behind our framework is that if the employed TSC models accurately classify \emph{stable} and \emph{unstable} measurements, then the framework will automatically halt a microbenchmark execution immediately after the warm-up phase ends, thereby providing high-quality performance results with the minimal required testing time.

%% file: design.tex
\section{Experimental Setup}\label{sec:design}

In this section, we outline the experimental procedure, describe the dataset and evaluation metrics employed in our study, and detail the implementation of our framework.


\subsection{Dataset of JMH Performance Measurements}

\input{tables/projects}

We utilize the dataset by Traini \etal \cite{Traini2022a}, which includes performance measurements from 586 JMH microbenchmarks across 30 Java software systems (refer to Table~\ref{tab:bench_proj} for an overview of these systems). To our knowledge, this is the most extensive publicly available dataset of JMH performance measurements.
The dataset includes 10 time series of performance measurements per microbenchmark, resulting in a total of 5,860 time series. Each time series comprises performance measurements gathered from 3,000 consecutive microbenchmark iterations performed within a fresh JVM instantiation (often referred to as a \emph{fork} in JMH nomenclature). Each data point in the time series reports the average execution time observed within the microbenchmark iteration.
In addition, each time series is annotated with the $st$ iteration number at which a steady-state was attained, based on the Barrett \etal's technique \cite{Barrett2017a}.

Note that this high number of JMH iterations would be impractical in a real-world scenario due to the extensive testing time required (\eg the entire data collection required about 93 days according to Traini \etal \cite{Traini2022a}). Nonetheless, we can leverage these time series along with the associated annotation $st$ to build a solid ground-truth for training our TSC models and validate the effectiveness of our framework.

\subsection{Experimental procedure}

\mysubsubsection{Dataset Preprocessing}
The initial phase of our experimental procedure involves generating the dataset for training and evaluating the TSC models. To ensure the effectiveness of the learning process, time series that do not reach a steady-state are excluded, leading to the omission of 10.9\% of the time series.

Our dataset construction is based on three main criteria: (i) ensure equal representation of segments from different time series, so no single series dominates the dataset, (ii) achieve a reasonable balance between \emph{stable} and \emph{unstable} segments, and (iii) ensure an adequate coverage of the time series while minimizing measurement redundancy.

To meet the first criterion, we sample 100 segments of 100 measurements each from each time series, following a measurement window similar to that used in prior work \cite{Laaber2020a}. This ensures equal representation of different time series segments in the dataset.
To satisfy the second criterion, we sample 50 \emph{stable} and 50 \emph{unstable} segments from each time series, ensuring a balance between the two classes within the dataset.
For the third criterion, we use a time series segmentation with adaptive step size to limit measurement redundancy across segments. This strategy is applied independently to both the warm-up and steady-state phases of the time series. Specifically, the segments are selected equidistantly, ensuring that the starting points of the segments are evenly spaced. Specifically, the window segmentation for the 50 \emph{unstable} segments during the warm-up phase uses an adaptive step size of 
\(step = \lfloor (st - 1) / 50 \rfloor \)
where $st$ denotes the iteration where the microbenchmark reaches the steady-state. Similarly, for the steady-state phase, the step size is computed as
\(step = \lfloor (n - st) / 50 \rfloor \), where $n$ denote the length of the time series, namely 3,000.
This approach ensures reasonable coverage of the time series while reducing potential measurement redundancy due to overlapping segments.

As a result, we obtain a final dataset of 521,900 measurement segments, including 376,925 (72\%) \emph{stable} and 144,975 (28\%) \emph{unstable} segments.

%
%
%

\mysubsubsection{Model Training} We adopt a 5-fold cross-validation process for each of the three TSC models. The dataset is divided into five folds of equal size. For each fold, the model is trained on four folds and tested on the remaining one. This process is repeated 5 times, with each fold being used exactly once as the test set. To prevent data leakage, we ensure that measurement segments gathered from the same microbenchmark always appear within the same fold. This guarantees an unbiased evaluation setup, where the model is tested against measurements gathered from unseen microbenchmarks.
Furthermore, to increase the representativeness and heterogeneity of each fold, we ensure that 
each fold has a similar proportion of microbenchmarks per project, using stratified random sampling.
The entire cross-validation training process for all the three TSC models required approximately 2 days to complete.

\mysubsubsection{Application} To evaluate our framework, we employ a methodology similar to prior work \cite{Laaber2020a}. Specifically, we mimic the application of the framework during microbenchmark execution through post-hoc analysis. We use a sliding measurement window of 100 performance measurements on the 5,860 time series of measurements from the dataset of JMH performance measurements \cite{Traini2022a}. For a given time series (\ie microbenchmark fork), the process begins with a segment containing the first consecutive 100 performance measurements. The framework submits this segment to the TSC model to determine if the measurements are classified as \emph{stable} or not. If the TSC model classifies the measurements as \emph{stable}, then the framework stops the process and returns the measurements $M$ for performance evaluation. Conversely, if the model deems the measurements \emph{unstable}, then the framework shifts the sliding window by one iteration, simulating the execution of an additional warm-up iteration. This process is repeated until the TSC model classifies the measurements as \emph{stable}.
Similar to prior work \cite{Laaber2020a}, we set an upper limit of 500 warm-up iterations. If this limit is reached, the process stops automatically, and the current set of measurements $M$ is returned for evaluation.

As a result of this process, for each microbenchmark fork, we obtain the number of warm-up iterations estimated by the framework, along with the corresponding set of measurements $M$ provided for performance evaluation. 

Similar to the model training phase, when evaluating the framework on a microbenchmark, we consistently use the model trained on the four folds that exclude the target microbenchmark. This approach ensures that our framework is always evaluated against a previously unseen microbenchmark.

\subsection{Evaluation metrics}

We use different metrics to evaluate the prediction accuracy of TSC models:

\begin{itemize}
\item \textbf{Precision} represents the ratio of true positive outcomes to the total positive predictions made by the model. In our context, the positives are \emph{stable} measurement segments.
\item \textbf{Recall} indicates the ratio of true positive outcomes to the total actual positives in the dataset.
\item \textbf{F1-score} is the harmonic mean of precision and recall, providing a single measure that balances both concerns.
\item \textbf{Balanced Accuracy} is the average recall obtained across each of the two classes. We use this metric instead of traditional accuracy due to the imbalanced nature of our dataset.
\end{itemize}

The following metrics are used for evaluating the application of our framework:

\begin{itemize}
	\item \textbf{Warm-up Estimation Error (WEE)} measures the accuracy of the warm-up iterations estimated by the framework. Specifically, it represents the absolute difference, in seconds, between the actual steady-state ($st$) and the end of the warm-up phase as estimated by our framework. This metric helps us understand how closely the warm-up iterations provided by our framework align with the actual warm-up phase of the microbenchmark.

	\item \textbf{Measurement Deviation} evaluates the quality of the performance test results provided by the framework. For a given microbenchmark, this metric compares the set of performance measurements $\mathcal{M}$ returned by the framework to the ground-truth steady-state measurements $\mathcal{M}^*$. The set $\mathcal{M}^*$ consists of all measurements gathered from the steady-state iteration $st$ onwards, for every time series of a particular microbenchmark\footnote{Time series where the steady-state is not reached are excluded.}. A large deviation between $\mathcal{M}$ and $\mathcal{M}^*$ indicates poor result quality.
To assess this deviation, we adhere to performance engineering best practices \cite{Traini2021a, Jangali2023a, Zhang2023a, Laaber2020a} by calculating the confidence interval for the execution time ratio~\cite{Kalibera2013a}. Specifically, we employ the bootstrap method \cite{Kalibera2012a, Davison1997} with 10,000 iterations \cite{Hesterberg2015}, using hierarchical random resampling with replacement at two levels \cite{Kalibera2012a}, and a significance level of $\alpha$ = 0.05.
If the confidence interval for the ratio includes 1, there is no statistically significant difference between $\mathcal{M}$ and $\mathcal{M}^*$. Conversely, if the interval does not include 1, it suggests that $\mathcal{M}$ does not reflect the observed steady-state performance and could therefore mislead performance evaluation.
For instance, a confidence interval of $(1.04, 1.06)$ indicates that $\mathcal{M}$ statistically differ from $\mathcal{M}^*$ by $5\%\pm1\%$ with 95\% confidence.

	\item \textbf{Testing Time} represents the total duration, in seconds, required to execute the microbenchmark using our framework. This duration includes both the warm-up iterations and the subsequent iterations used to collect performance measurements.
\end{itemize}



\subsection{Implementation details}

To facilitate the convergence of TSC models, we standardize each performance measurement $x$ within each segment of the training dataset using the formula $(x - \mu)/\sigma$, where $\mu$ is the segment mean and $\sigma$ is the segment standard deviation.
We implement each of the TSC model as follows:
\begin{itemize}
\item \textbf{FCN}: We reimplement the neural network using TensorFlow\footnote{\url{https://www.tensorflow.org}}. Our implementation follows the hyperparameters specified in the original paper \cite{Wang2017a}. Specifically, we use three 1-D kernels with sizes \{8, 5, 3\} and three convolution blocks with filter sizes \{128, 256, 128\}. 
\item \textbf{OSCNN}: We reuse the original implementation provided in the replication package of Tang \etal \cite{Tang2022a}, which uses PyTorch\footnote{\url{https://pytorch.org}}. We maintained the default hyperparameter settings defined by the authors.
\item \textbf{ROCKET}: We leverage the implementation provided by aeon\footnote{\url{https://www.aeon-toolkit.org}}. We set the number of kernels to 500.
\end{itemize}

When training the neural networks (\ie FCN and OSCNN), we set aside 25\% of the training set as a validation set. Both FCN and OSCNN are trained using the Adam optimizer \cite{Kingma2015} with a learning rate of 0.001, $\beta_{1}$=0.9, $\beta_{2}$=0.999, and $\epsilon$=1e$^{-8}$. If the validation loss does not improve after 20 epochs, we halve the learning rate.
We train the neural networks for up to 500 epochs, employing an early stopping strategy that halts training if the validation loss does not improve for 50 consecutive epochs. Throughout the training process, we save the model weights that yield the best validation loss. These optimized weights are then used for evaluation.

Since both FCN and OSCNN return probabilities rather than class labels, we tune the decision threshold to optimize Youden's index \cite{Youden1950} on the validation set, rather than relying on a traditional 0.5 threshold. This tuning is not applicable to ROCKET, as it directly returns class labels.

The ROCKET training process does not use an explicit validation set, however, the implementation provided by aeon internally utilizes leave-one-out cross-validation for training the Ridge classifier.

For all the TSC models, we use a batch size of 1,024.
 

%% file: tables/projects.tex
\begin{table}[t]
\footnotesize
\center
\caption{ Overview of the Java systems, including the name of each GitHub repository, the number of stars and forked repositories, and the number of microbenchmarks involved in the dataset.}
\label{tab:bench_proj}
\begin{tabular}{lrrr}
  \toprule
         \textbf{Repository} & \textbf{Stars} & \textbf{Forked Repo.} & \textbf{Microbench.}\\
  \midrule
      HdrHistogram/HdrHistogram & 2,141 & 251 & 20 \\
                JCTools/JCTools & 3,496 & 554 & 20 \\
               ReactiveX/RxJava & 47,702 & 7,581 & 20 \\
    RoaringBitmap/RoaringBitmap & 3,415 & 535 & 20 \\
                   apache/arrow & 13,686 & 3,341 & 20 \\
                   apache/camel & 5,366 & 4,896 & 20 \\
                    apache/hive & 5,370 & 4,601 & 20 \\
                   apache/kafka & 27,594 & 13,571 & 20 \\
          apache/logging-log4j2 & 3,290 & 1,559 & 20 \\
               apache/tinkerpop & 1,911 & 786 & 20 \\
  cantaloupe-project/cantaloupe & 261 & 104 & 19 \\
                    crate/crate & 3,977 & 546 & 20 \\
           eclipse-vertx/vert.x & 14,153 & 2,043 & 16 \\
    eclipse/eclipse-collections & 2,368 & 581 & 20 \\
          eclipse/jetty.project & 3,766 & 1,899 & 19 \\
                  eclipse/rdf4j & 347 & 160 & 20 \\
                    h2oai/h2o-3 & 6,756 & 1,991 & 20 \\
            hazelcast/hazelcast & 5,935 & 1,803 & 17 \\
                 imglib/imglib2 & 291 & 93 & 20 \\
                      jdbi/jdbi & 1,917 & 333 & 15 \\
                jgrapht/jgrapht & 2,535 & 819 & 20 \\
                    netty/netty & 32,923 & 15,763 & 20 \\
              openzipkin/zipkin & 16,780 & 3,069 & 20 \\
                prestodb/presto & 15,646 & 5,265 & 20 \\
         prometheus/client\_java & 2,134 & 772 & 20 \\
          protostuff/protostuff & 2,016 & 302 & 20 \\
                 r2dbc/r2dbc-h2 & 196 & 44 & 20 \\
               raphw/byte-buddy & 6,052 & 776 & 20 \\
      yellowstonegames/SquidLib & 447 & 46 & 20 \\
                zalando/logbook & 1,737 & 257 & 20 \\
  \bottomrule
  \end{tabular}

\end{table}

%% file: results.tex
\section{Results}\label{sec:results}
In this section, we discuss the study results by posing and answering three research questions.

\subsection{RQ$_1$: To what extent can TSC models accurately classify stable and unstable measurements?}

\mysubsubsection{Motivation}
We aim to evaluate the capabilities of TSC models in classifying measurements gathered from the steady-state and warm-up phases of microbenchmark execution. These results provide initial insights into the potential suitability of TSC models for dynamically halting warm-up iterations.

\mysubsubsection{Approach} For each of the three TSC models, we compute the average precision, recall, F1-score and balanced accuracy across the five folds. 

\input{tables/rq1}
\mysubsubsection{Results}
\tabref{tab:rq1} shows the results of the TSC models. Overall, we find that TSC models demonstrate good prediction accuracy in classifying \emph{stable} and \emph{unstable} segments, with F1-scores ranging from 0.748 to 0.867 and balanced accuracy between 0.682 and 0.712.
ROCKET stands out as the best-performing model in terms of F1-score, while neural network models achieve higher balanced accuracy. The lower F1-scores of FCN and OSCNN are primarily due to their lower recall rates, which are 0.659 and 0.65, respectively. Conversely, ROCKET shows a notably high recall on stable segments (0.932) but has a balanced accuracy of only 0.682, which indicates a limited recall on \emph{unstable} segments ($\sim$0.43).
This behavior can be attributed to the higher number of false positives predicted by ROCKET, which is also reflected in its lower precision scores compared to FCN and OSCNN (0.81 \emph{vs.} 0.88 and 0.886). 
It is important to note that false positives can be quite detrimental to our framework, as they may erroneously stop the microbenchmark execution, with no way to recover from the false prediction. In contrast, false negatives are less problematic since they do not halt execution, thus giving the framework another chance to correctly classify the stable measurements in the subsequent iteration.
Nonetheless, in the following research questions, we will examine how the prediction accuracy of different models influences the overall effectiveness of our framework.

\begin{tcolorbox}[size=title]
	\textbf{Answer to RQ$_1$:} 
TSC models effectively classify \emph{stable} and \emph{unstable} measurements, so demonstrating their suitability for dynamically halting warm-up iterations. This supports their integration into our framework.
\end{tcolorbox}

\subsection{RQ$_2$: How does our AI-based framework compare to the state-of-practice (SOP) in Java microbenchmarking?}

\mysubsubsection{Motivation}
Developers typically use a fixed number of warm-up iterations that are defined beforehand based on their domain expertise.
With this research question, we aim to understand if the use of our framework for dynamically halting warm-up iterations provides advantages over the SOP.

\mysubsubsection{Approach} We extract the number of warm-up iterations specified by the developers for each microbenchmark using a method similar to \cite{Traini2022a, Imran2024}. We then compare the WEE of our framework with that of the SOP for each time series that reaches a steady-state.
For this comparison, we employ the Wilcoxon signed-rank test \cite{Wilcoxon1945} along with two measures of effect size: the Vargha-Delaney \vda \cite{Vargha2000} and the matched pairs rank biserial correlation  $r$  \cite{Kerby2014}. In our context, the \vda measures the proportion of pairs where the WEE of our framework is lower than that of the developers.  An \vda value $>0.5$ indicates that our framework provides a more accurate estimation of the warm-up phase than the SOP.
The matched pairs rank biserial correlation $r$ represents the difference between the proportion of favorable and unfavorable evidence; in our case, favorable evidence indicates a lower WEE for the framework. Thus, an  $r>0$ indicates that our framework performs better than the SOP.

In addition to evaluating the accuracy of the warm-up estimation, we examine how the adoption of the framework affects the quality of performance test results and the testing time for each microbenchmark.
To evaluate the impact of our framework on result quality, we calculate the percentage of microbenchmarks that show an improvement (or regression) compared to the SOP.
For a given microbenchmark, an improvement in result quality occurs under the following two conditions: (i) the performance measurements from SOP ($\mathcal{M}_{SOP}$) are statistically different from those of the steady-state ($\mathcal{M}^*$) (\ie the confidence interval for the execution time ratio does not include 1), and (ii) the performance measurements provided by the framework $\mathcal{M}$ are not statistically different from $\mathcal{M}^*$ (\ie the confidence interval includes 1). This indicates that the framework improves result quality, by providing measurements that more faithfully represent the microbenchmark true steady-state performance.
A result quality regression indicates the opposite situation.

We also report the percentage of microbenchmarks where the framework shows improvement (or regression) in terms of testing time.
A microbenchmark is considered improved if the framework reduces its testing time when compared to SOP.
Note that we only consider testing time improvements (or regressions) in cases where the measurements provided by both the framework and the SOP are not statistically different from the steady-state. This ensures that a reduction in testing time is not mistakenly interpreted as an improvement if it leads to poor result quality.

To achieve a fair comparison, we derive the set of performance measurements $\mathcal{M}$ returned by the framework, as well as the microbenchmark testing time, using the same number of measurement iterations and JMH forks specified by the developers (more details on this process are available in our replication package).


\input{tables/rq2_wee}
\input{tables/rq2_impr}
\mysubsubsection{Results}
We discuss the results of this RQ from different aspects.

\subsubsubsection{Warm-up estimation accuracy}
\tabref{tab:rq2_wee} shows the results of the comparison between the WEE of the framework and the SOP. We observe that the framework delivers more accurate estimates of the warm-up phase across all employed TSC models. The results of the Wilcoxon test show that the differences are statistically significant (p $<$ 0.001) for all models, with medium effect sizes (\vda ranging from 0.658 to 0.683). Additionally, the matched pairs rank biserial correlation results emphasize a considerable gap between the favorable and unfavorable outcomes, with positive $r$ values ranging from 0.282 to 0.352. This indicates a higher likelihood of achieving lower WEE with our framework.

\subsubsubsection{Result quality}
Table \ref{tab:rq2_impr} presents the percentages of improvements and regressions in result quality and testing time achieved by our framework compared to the SOP. We observe that our framework enhances result quality in 16.7\%, 17.9\%, and 9.4\% of the microbenchmarks when using FCN, OSCNN, and ROCKET, respectively.  However, it also exhibits a number of regressions, ranging from 13.7\% to 25.9\%.

ROCKET yields the worst result quality among the TSC models, with only 9.4\% improvements and 25.9\% regressions. This poor result quality is likely due to its tendency to underestimate the warm-up phase, thus leading to the collection of measurements that significantly deviate from the true steady-state performance. As illustrated in \figref{fig:res_warmup_estimation}, ROCKET underestimates the warm-up phase in 70.3\% of the time series, while the SOP only does so in 36\% of cases. This propensity to stop the microbenchmark earlier could be linked to the higher number of false positives produced by ROCKET (see RQ$_1$ discussion).

In contrast, OSCNN provides the best result quality among the TSC models, showing a slight tendency towards improvement compared to SOP, with 17.9\% improvements and 13.7\% regressions. Additionally, the degree of measurement deviation is lower in OSCNN, as evident from \tabref{tab:rq23_pd_tt}. This table presents descriptive statistics of the relative measurement deviation, which we use to quantify the magnitude of deviation from the steady-state. This metric is calculated as the absolute value of the difference between the center of the confidence interval for the execution time ratio and one. 
The table shows that the measurements returned by OSCNN deviate less from the steady-state than those of SOP, with a median deviation of 5.5\% (\emph{vs.} 6.5\% in SOP) and an interquartile range (IQR) of 2.4-12.3\% (\emph{vs.} 2.4-19.2\%).

\subsubsubsection{Testing time}
From \tabref{tab:rq2_impr}, we observe that our framework reduces the testing time of SOP across a large percentage of microbenchmarks, showing improvements by 30.7\%, 30.9\%, and 20.1\% for FCN, OSCNN, and ROCKET, respectively. 
By looking at \tabref{tab:rq23_pd_tt}, we also note that using the framework significantly reduces the median testing time of the SOP by 21\% to 49\%.  For instance, FCN shows a median (IQR) testing time of 74 (43-150) seconds, while the SOP provides a median testing time of 100 (30-403) seconds.
These findings indicate that, for a significant portion of microbenchmarks, our framework drastically reduces the testing time of the SOP without compromising the result quality. This enhancement can be attributed to the higher accuracy achieved by the framework in estimating the end of the warm-up phase.  Indeed, as shown in \figref{fig:res_warmup_estimation}, SOP overestimates the end of the warm-up phase in 63.9\% of cases, thereby increasing the testing time. In contrast, our framework overestimates it in only 38.1\% (FCN), 41.2\% (OSCNN), and 13\% (ROCKET) of the cases, respectively. Moreover, we observe that our framework can exactly identify the end of the warm-up phase in 15.5\%, 14.9\%, and 16.6\% of the cases, respectively, whereas the SOP does so in only 0.1\% of the cases.

\subsubsubsection{Overall} We find that the framework improves either the result quality or the testing time of the SOP in 47.4\%, 48.8\%, and 29.5\% of the microbenchmarks, depending on the TSC model employed (see \tabref{tab:rq2_impr}). Conversely, it shows regressions in 22.2\%, 21.8\%, and 32.4\% of the cases. This yields a net improvement of +25.3\% for FCN and +27\% for OSCNN, and a net regression of -2.9\% for ROCKET.

\begin{tcolorbox}[size=title]
	\textbf{Answer to RQ$_2$:}
Our framework provides more accurate estimates of the warm-up phase compared to the SOP. This higher accuracy translates into a net improvement in either result quality or testing time in up to +27\% of the microbenchmarks, with OSCNN demonstrating the highest net improvement.
\end{tcolorbox}

\input{tables/rq23_pd_tt}

\subsection{RQ$_3$: How does our AI-based framework compare to the state-of-the-art (SOTA) in Java microbenchmarking?}

\noindent\textbf{Motivation.}
This research question aims to compare our framework against prior SOTA approaches that dynamically halt warm-up iterations. 
Our objective is to assess whether the use of sophisticated TSC models provides benefits over traditional dynamic approaches.

\medskip\noindent\textbf{Approach.} We compare the WEE of our framework to the dynamic technique proposed by Laaber \etal \cite{Laaber2020a}, by evaluating all three variants of their technique: COV, RCIW, and KLD. To determine the warm-up iterations for each variant, we use the original replication package provided by the authors \cite{Laaber2020a}. As in RQ$_2$, we employ the Wilcoxon signed-rank test \cite{Wilcoxon1945}, the Vargha-Delaney \vda, and the matched pairs rank biserial correlation $r$ to assess whether our framework provides more accurate estimates of the warm-up phase.

Additionally, we report the percentages of improvements and regressions in both results quality and testing time. For an unbiased comparison, we derive the set of performance measurements, $\mathcal{M}$, and the testing time for our framework using the same measurement window defined by Laaber \etal (\ie 100 measurement iterations in our setup) and the same number of JMH forks (the number of forks is dynamically estimated for each microbenchmark in Laaber \etal's technique).

\medskip\noindent\textbf{Results.}
Similarly to RQ2, we discuss the results of this RQ from various perspectives.

\subsubsubsection{Warm-up estimation accuracy}
\tabref{tab:rq3_wee} shows the results of the comparison between the WEE provided by the framework and those of SOTA. We find that the framework provides more accurate estimations of the warm-up phase across all the models than all SOTA variants ones. The differences are statistically significant (p $<$ 0.001) for all comparisons,  based on the results of the Wilcoxon test.  The \vda values range from 0.621 to 0.731, indicating a small to large effect size according to the thresholds defined by Vargha and Delaney \cite{Vargha2000}. The rank biserial correlation $r$ ranges from 0.157 to 0.414, suggesting a prevalence for the favorable outcome.

\input{tables/rq3_wee} 

\input{tables/rq3_impr}

\subsubsubsection{Result quality}
\tabref{tab:rq3_impr} presents the percentages of improvement and regression in result quality and testing time. We observe that ROCKET performs worse than SOTA approaches in terms of result quality. Specifically, ROCKET produces regressions in 24.6\%, 51.4\%, and 22.5\% of the microbenchmarks when compared to CV, RCIW, and KLD, respectively. Moreover, it reports improvements in only 11.3\%, 4.1\%, and 7.2\% of the cases. This limitation might once again be attributed to the higher false positive rate produced by this model.
In contrast, we observe that neural network models generally perform better than SOTA in terms of result quality, with the only exception being RCIW. For instance, OSCNN improves the result quality of CV and KLD in 28.8\% and 27.8\% of microbenchmarks, respectively, and causes regressions in 12.3\% and 12.1\% of them. Moreover, OSCNN produces lower relative measurement deviations than CV and KLD, as shown in \tabref{tab:rq23_pd_tt}. The median (IQR) deviation for OSCNN is 5.5\% (2.4-12.3\%), while CV and KLD induce deviations of 9.9\% (4.7-17.5\%) and 7.9\% (4.2-13.4\%), respectively. We observe a similar trend of improvements in FCN, albeit with slightly less pronounced results.

The framework provides lower result quality than RCIW across all TSC models, causing regressions in 24.2\% (FCN), 21.8\% (OSCNN), and 51.4\% (ROCKET) of the microbenchmarks. This behavior can be attributed to the RCIW's higher tendency to overestimate the warm-up phase, observed in 74.1\% of the cases (see \figref{fig:res_warmup_estimation}), which increases the chance of gathering steady-state measurements. While this tendency ensures better result quality compared to our framework, it also leads to longer testing times.

\subsubsubsection{Testing time}
As shown in \tabref{tab:rq23_pd_tt}, RCIW reports a median (IQR) testing time of 300 (265-301) seconds, while the most time-consuming TSC model of our framework, OSCNN, produces a median (IQR) testing time of 79 (47-161) seconds, which is approximately one-quarter of RCIW testing time. Furthermore, from \tabref{tab:rq3_impr}, we observe that FCN and OSCNN can improve the testing time for about half of the microbenchmarks (50.3\% and 52.6\%, respectively) without affecting result quality.
When compared to the other two SOTA variants, CV and KLD, our framework still shows improvements in testing time. For example, OSCNN reduces the testing time in 26.8\% and 23\% of the microbenchmarks, respectively.

\subsubsubsection{Overall}
We find that, when employing neural network models such as FCN and OSCNN, our framework provides improvements over the SOTA in either result quality or testing time in approximately half of the microbenchmarks, with percentages ranging from 50.3\% to 59.6\%. The percentages of regressions are lower, with values ranging from 20\% to 28.8\%, thus resulting in substantial net improvements. For instance, from \tabref{tab:rq3_impr}, we can observe that OSCNN provides a net improvement over CV, RCIW, and KLD by +35.3\%, +32.9\%, and +25.9\%, respectively.


\begin{tcolorbox}[size=title]
	\textbf{Answer to RQ$_3$:}
Our framework provides more accurate estimates of the warm-up phase than the SOTA. While ROCKET often reduces result quality, variants of the framework based on neural network models observably enhance either the result quality or testing time of the SOTA techniques, leading to net improvements in up to +35.3\% of the microbenchmarks. 
\end{tcolorbox}

%% file: tables/rq1.tex
\begin{table}[t]
\small
\caption{Results for the classification of segments (RQ$_1$).}
\label{tab:rq1}
\center
\begin{tabular}{lcccc}
\toprule
\textbf{Model} & \textbf{Prec.} & \textbf{Rec.} & \textbf{F1} & \textbf{Bal. Acc.} \\
\midrule
FCN & 0.880 & 0.659 & 0.753 & 0.712 \\
OSCNN & 0.886 & 0.650 & 0.748 & 0.715 \\
ROCKET & 0.810 & 0.932 & 0.867 & 0.682 \\
\bottomrule
\end{tabular}
\end{table}

%% file: tables/rq2_wee.tex
\begin{table}[t]
\small
\center
\caption{Results for the WEE comparison of models \emph{vs.} SOP (RQ$_2$).}
\label{tab:rq2_wee}
\begin{tabular}{lccc}
\toprule
\textbf{Model \emph{vs.} SOP} & $p$-value & \vda & $r$ \\
\midrule
FCN \emph{vs.} SOP & $<$0.001 & 0.664 & 0.352 \\
OSCNN \emph{vs.} SOP & $<$0.001 & 0.658 & 0.348 \\
ROCKET \emph{vs.} SOP & $<$0.001 & 0.683 & 0.282 \\
\bottomrule
\end{tabular}
\end{table}

%% file: tables/rq2_impr.tex
\begin{table*}[t]
\small
\center
\caption{Percentages of improvement and regression when comparing models \emph{vs.} SOP (RQ$_2$).}
\label{tab:rq2_impr}
\begin{tabular}{l|ccc|ccc|c}
\toprule
 & \multicolumn{3}{c}{\textbf{Improvement (\%)}} & \multicolumn{3}{c}{\textbf{Regression (\%)}} & \textbf{Net Improvement (\%)} \\
\textbf{Model \emph{vs.} SOP} & Res. Quality & Testing Time & \textbf{\textit{Total}} & Res. Quality & Testing Time & \textbf{\textit{Total}} & (\textbf{\textit{Tot. Impr. - Tot. Regr.}}) \\
\midrule
FCN  \emph{vs.} SOP & 16.7 & 30.7 & \textbf{\textit{47.4}} & 14.3 & 7.8 & \textbf{\textit{22.2}} & \textbf{\textit{+25.3}} \\
OSCNN  \emph{vs.} SOP & 17.9 & 30.9 & \textbf{\textit{48.8}} & 13.7 & 8.2 & \textbf{\textit{21.8}} & \textbf{\textit{+27.0}} \\
Rocket  \emph{vs.} SOP & 9.4 & 20.1 & \textbf{\textit{29.5}} & 25.9 & 6.5 & \textbf{\textit{32.4}} & \textbf{\textit{-2.9}} \\
\bottomrule
\end{tabular}
\end{table*}

\begin{figure*}[t]
\center
\includegraphics[width=0.8\linewidth]{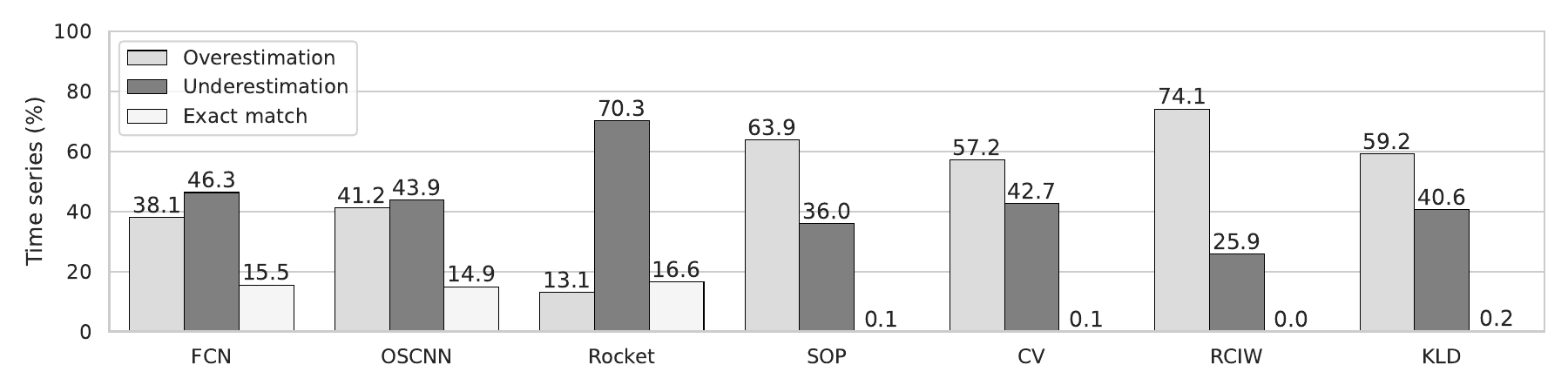}
\caption{Percentages of \emph{overestimation}, \emph{underestimation}, and \emph{exact match} for models/SOP/SOTA (RQ$_{2/3}$).}
\label{fig:res_warmup_estimation}
\end{figure*}

%% file: tables/rq23_pd_tt.tex
\begin{table}[t]
\small
\center
\caption{Relative measurement deviation and testing time of models, SOP, and SOTA (RQ$_{2/3}$).}
\label{tab:rq23_pd_tt}
\begin{tabular}{lrr}
\toprule
 & \textbf{Rel. Meas. Dev. (\%)} & \textbf{Testing Time (\textit{sec})} \\
\textbf{Model/SOP/SOTA} & Median (IQR)  & Median (IQR) \\
\midrule
FCN & 5.7 (2.8-11.3) & 74 (43-150) \\
OSCNN & 5.5 (2.4-12.3) & 79 (47-161) \\
Rocket & 10.0 (4.5-21.7) & 51 (30-78) \\
\midrule
SOP & 6.5 (2.4-19.2) & 100 (30-403) \\
\midrule
CV & 9.9 (4.7-17.5) & 75 (51-113) \\
RCIW & 4.5 (1.8-11.5) & 300 (265-301) \\
KLD & 7.9 (4.2-13.4) & 121 (94-156) \\
\bottomrule
\end{tabular}
\end{table}

%% file: tables/rq3_wee.tex
\begin{table}[t]
\small
\center
\caption{Results for the WEE comparison of models \emph{vs.} SOTA (RQ$_3$).}
\label{tab:rq3_wee}
\begin{tabular}{lccc}
\toprule
\textbf{Model \emph{vs.} SOTA} & $p$-value & \vda & $r$ \\
\midrule
FCN \emph{vs.} CV & $<$0.001 & 0.637 & 0.370 \\
FCN \emph{vs.} RCIW & $<$0.001 & 0.731 & 0.400 \\
FCN \emph{vs.} KLD & $<$0.001 & 0.630 & 0.292 \\
\midrule
OSCNN \emph{vs.} CV & $<$0.001 & 0.632 & 0.385 \\
OSCNN \emph{vs.} RCIW & $<$0.001 & 0.728 & 0.414 \\
OSCNN \emph{vs.} KLD & $<$0.001 & 0.621 & 0.288 \\
\midrule
ROCKET \emph{vs.} CV & $<$0.001 & 0.649 & 0.245 \\
ROCKET \emph{vs.} RCIW & $<$0.001 & 0.726 & 0.264 \\
ROCKET \emph{vs.} KLD & $<$0.001 & 0.656 & 0.157 \\
\bottomrule
\end{tabular}
\end{table}

%% file: tables/rq3_impr.tex
\begin{table*}[ht]
\small
\center
\caption{Percentages of improvement and regression when comparing models \emph{vs.} SOTA (RQ$_3$).}\label{tab:rq3_impr}
\begin{tabular}{l|ccc|ccc|c}
\toprule
 & \multicolumn{3}{c}{\textbf{Improvement (\%)}} & \multicolumn{3}{c}{\textbf{Regression (\%)}} & \textbf{Net Improvement (\%)} \\
\textbf{Model \emph{vs.} SOTA} & Res. Quality & Testing Time & \textbf{\textit{Total}} & Res. Quality & Testing Time & \textbf{\textit{Total}} & (\textbf{\textit{Tot. Impr. - Tot. Regr.}}) \\
\midrule
FCN  \emph{vs.} CV & 26.8 & 27.1 & \textbf{\textit{53.9}} & 12.3 & 7.7 & \textbf{\textit{20.0}} & \textbf{\textit{+34.0}} \\
FCN  \emph{vs.} RCIW & 6.3 & 50.3 & \textbf{\textit{56.7}} & 24.2 & 4.6 & \textbf{\textit{28.8}} & \textbf{\textit{+27.8}} \\
FCN  \emph{vs.} KLD & 25.9 & 24.4 & \textbf{\textit{50.3}} & 13.1 & 10.4 & \textbf{\textit{23.5}} & \textbf{\textit{+26.8}} \\
\midrule
OSCNN  \emph{vs.} CV & 28.8 & 26.8 & \textbf{\textit{55.6}} & 12.3 & 8.0 & \textbf{\textit{20.3}} & \textbf{\textit{+35.3}} \\
OSCNN  \emph{vs.} RCIW & 7.0 & 52.6 & \textbf{\textit{59.6}} & 21.8 & 4.8 & \textbf{\textit{26.6}} & \textbf{\textit{+32.9}} \\
OSCNN  \emph{vs.} KLD & 27.8 & 23.0 & \textbf{\textit{50.9}} & 12.1 & 12.8 & \textbf{\textit{24.9}} & \textbf{\textit{+25.9}} \\
\midrule
ROCKET  \emph{vs.} CV & 11.3 & 19.8 & \textbf{\textit{31.1}} & 24.6 & 2.7 & \textbf{\textit{27.3}} & \textbf{\textit{+3.8}} \\
ROCKET  \emph{vs.} RCIW & 4.1 & 24.4 & \textbf{\textit{28.5}} & 51.4 & 3.4 & \textbf{\textit{54.8}} & \textbf{\textit{-26.3}} \\
ROCKET  \emph{vs.} KLD & 7.2 & 21.5 & \textbf{\textit{28.7}} & 22.5 & 3.9 & \textbf{\textit{26.5}} & \textbf{\textit{+2.2}} \\
\bottomrule
\end{tabular}
\end{table*}

%% file: threats.tex
\section{Threats to validity}\label{sec:threats}
\smallskip
\textbf{Construct validity.}
We derive the number of warm-up iterations through post-hoc analysis using a methodology similar to \cite{Laaber2020a, Traini2022a}.
This evaluation methodology does not account for the overhead introduced by the TSC model prediction. However, we do not consider this overhead for both our framework and the SOTA techniques when calculating the testing time, whereas SOP does not involve any overhead, thus ensuring a fair comparison.
Moreover, to further mitigate this threat, we measured the inference time of each TSC model on a sample of 500 measurement segments, and we obtained a median overhead per microbenchmark iteration of 9\% for FCN, 2\% for OSCNN, and 2\% for ROCKET.
These overheads appear minimal when compared to the testing time differences observed in \tabref{tab:rq23_pd_tt}. Therefore, they are unlikely to impact our main findings.

We rely on the notion of steady-state as defined by Barrett~\etal~\cite{Barrett2017a}, whereas using an alternative notion may change the study outcomes. 
We integrate three state-of-the-art TSC models into our framework, using different models may yield different results.

The proposed framework aims to dynamically infer the appropriate number of warm-up iterations at runtime. Although other relevant parameters, such as the number of forks and measurement iterations, can influence the quality of results and the testing time of microbenchmarks, these aspects are beyond the scope of our work. Nonetheless, to ensure fairness in our evaluation, we consistently use the same number of forks and measurement iterations when comparing our approach with baselines.



\mysubsubsection{Internal validity}
TSC model training involves randomness, hence there might be slight differences in the results when re-executing the experiments.
We employ TSC models by using default hyper-parameters, whereas using different hyper-parameters may change the experiments outcomes.

\mysubsubsection{External validity}
The findings of our study may not generalize to microbenchmarks beyond our experiment dataset. However, we evaluate our framework on 583 microbenchmarks from 30 well-established OSS projects spanning various domains (see \tabref{tab:bench_proj}). Furthermore, the number of systems involved in our study is larger than the ones considered in most recent software performance studies \cite{Jangali2023a, Ding2020, Chen2022a, Laaber2024,Laaber2020a}.

Each TSC model is trained on an average of 417,520 segments from 468 microbenchmarks per fold iteration. Using training sets of different sizes and heterogeneity could affect the prediction accuracy of the models, and consequently, the outcome of the framework. We encourage future dedicated analyses to investigate how these factors might affect the framework effectiveness.

We utilize the dataset of JMH measurements from our previous work \cite{Traini2022a}, which was collected using a rigorous procedure to minimize noise. We specifically choose these measurements to reduce confounding factors that could compromise the soundness of our study. However, using measurements from more variable environments (\eg cloud) could lead to different study outcomes.

%% file: related.tex
\section{Related Work}\label{sec:related}


Besides the works discussed in \secref{sec:background}, other techniques have been proposed to estimate the attainment of steady-state in performance tests. Kalibera and Jones \cite{Kalibera2013a} introduced a methodology that relies on the visual analysis of auto-correlation function plots, lag plots, and run-sequence plots. However, this technique requires manual analysis, making it impractical for real-world use cases.
Barrett \etal \cite{Barrett2017a} proposed an automated technique based on change point analysis \cite{Killick2012} to identify shifts in the time series of execution times and determine the attainment of the steady-state. While this technique is highly useful in scientific settings where a rigorous notion of steady-state is necessary, it is also impractical for real-world use due to the significant time effort required to run the performance tests.
AlGhamdi \etal \cite{AlGhamdi2023} proposed a technique to stop load tests when performance metric values become repetitive.
He \etal \cite{He2019a} proposed a statistical approach to halt performance tests in the cloud.
Abdullah \etal \cite{Abdullah2023} proposed an approach for the early stopping of non-productive experiments in performance testing.

%% file: conclusion.tex
\section{Conclusion}\label{sec:conclusion}

In this paper, we propose and study an AI-based framework to dynamically stop warm-up iterations in Java performance tests.
The results show that, when integrated with certain TSC models, our framework delivers result quality comparable to the state-of-practice while drastically reducing the testing time.
Furthermore, we find that the framework improves both the result quality and testing time in most of the investigated state-of-the-art techniques. 
This higher effectiveness can be attributed to the ability of TSC models to capture complex, non-linear patterns in measurements associated with steady-state execution, which simple statistical measures---such as those employed by state-of-the-art techniques--may overlook.

Our work has significant implications for both practitioners and researchers. Practitioners can use our framework to ensure result quality, while significantly reducing testing time of performance testing suites.
Researchers can integrate our framework into advanced software engineering techniques (such as, genetic improvement \cite{Langdon2015,Petke2018}, configuration tuning \cite{Chen2021b}, or software self-adaptation \cite{Wang2022a}), where rigorous and time-efficient performance evaluation is crucial. Additionally, researchers can leverage our framework to experiment with different (potentially more effective) TSC models, or use it as a robust baseline for future techniques that dynamically stop warm-up iterations at runtime.
Despite the promising results, our findings still highlight room for improvement in this area, particularly in the enhancement of the quality of performance test results. We encourage future research efforts aimed at this goal.
